\input epsf

\documentstyle[12pt,fleqn,titlepage]{article}
\newcommand{\dis}{\displaystyle}

\begin{document}
\bibliographystyle{unsrt}

\begin{titlepage} 
\par
\begin{flushright}
prepared for {\it Fluid Phase Equilibria} 
\end{flushright}

\vspace{2cm}

\par 
\begin{center} 
\begin{Large} 
Phase Separation in Polymer Solutions from a \\
Born-Green-Yvon Lattice Theory
\footnote{Paper presented at the Thirteenth 
Symposium on Thermophysical
Properties, June 22--27, 1997, Boulder, Colorado, U.S.A. } 
\end{Large} 
\par 
\vspace{1cm}
J. Luettmer-Strathmann and J.~E.~G. Lipson \\
Department of Chemistry \\ 
Dartmouth College,  Hanover, NH 03755 
\par 
\vspace{3cm} 
\end{center} 

\bigskip 
\begin{sloppypar} 
\end{sloppypar}

\end{titlepage} 

\pagestyle{plain} 
\section*{Abstract}

Phase separation in mixtures of polymers and alkanes is investigated
with the aid of a recently developed lattice model, based on the
Born-Green-Yvon (BGY) integral equation approach to fluids.
The system-dependent parameters of the BGY lattice model
for binary mixtures are deduced from those of its pure components, 
which in turn are  determined from a comparison with experimental data.
The lower critical solution temperatures (LCST's) for polyethylene
in various n-alkanes were predicted
from the BGY lattice model and compared with experimental data.
While the model underestimates the
LCST's for the smaller alkanes it reproduces the experimental values
very well for the higher alkanes (decane through tridecane).
The effect of the chain length of the polymer on the solubility
is investigated for the case of decane as a solvent.

\section*{Introduction}\label{intro}
In many polymer/solvent systems, phase separation occurs at a lower
critical solution temperature (LCST) \cite{fr65}.
For temperatures smaller
than the LCST the polymer is soluble while a miscibility gap opens
above the LCST which is closed again at an upper critical
solution temperature (UCST). These lower critical solution temperatures
are difficult to predict accurately from theory \cite{or67b,pa68,ko78}.
In this work, we employ a recently developed lattice 
model~\cite{li91,li92,li92b,li93}
based on the Born-Green-Yvon (BGY) integral equation approach to fluids to
investigate the miscibility of polyethylene in alkanes. The BGY lattice
model provides a compact expression for the configurational part
of the Helmholtz free-energy density which
allows the calculation of the relevant thermodynamic properties of
a mixture. 
In addition to the system-dependent parameters of the pure components, 
the lattice model for a binary mixture contains an energy parameter
describing interactions between unlike molecules.
In this work the 
interaction energy is approximated by the geometric mean
of the energy parameters of the pure components, allowing the thermodynamic
properties of the mixture to be predicted from those of the pure
components. In the following sections we describe how the BGY lattice
model is used to predict the LCST's and corresponding coexistence
curves for polyethylene in various alkanes, and then compare our results
with experimental data.

\section*{Born-Green-Yvon lattice model for fluids}\label{bgy}

In the BGY lattice model, a molecule of species $i$ is assumed to
occupy $r_i$ contiguous sites on a lattice having coordination number
$z$ and a total of $N_0$ lattice sites. As in earlier 
work~\cite{li92b,li93}, 
we set $z=6$. The site-fraction $\phi_i$
is defined as $\phi_i=r_iN_i/N_0$, where $N_i$ is the number of molecules
of component $i$ while $\phi_{\rm h}=N_{\rm h}/N_0$ is the fraction
of empty sites or holes, where $N_{\rm h}$ is the number of holes.
Each molecule of species $i$ has $q_iz=r_i(z-2)+2$ interaction sites,
leading to the definition of concentration variables
$\xi_i=q_iN_i/(N_{\rm h}+\sum_j q_jN_j)$ and
$\xi_{\rm h}=N_{\rm h}/(N_{\rm h}+\sum_j q_jN_j)$  which account for
the nearest-neighbor connectivity of the molecules.
The interaction energy associated with nonbonded nearest neighbors
of the same species is $\epsilon_{ii}$, while $\epsilon_{ij}$
corresponds to interactions between unlike nearest neighbor segments.
Using the Born-Green-Yvon integral
equation approach,  Lipson \cite{li91,li92}
derived an expression for the dimensionless configurational
Helmholtz free energy per lattice site $\hat{a}=\beta A/N_0$ of an
$M$-component mixture
\begin{equation}\label{bgy1}
\hat{a}=\frac{\dis \beta A}{\dis N_0}=\sum_{i={\rm h},1}^{M}\left(
\frac{\dis \phi_i}{\dis r_i}\ln{\phi_i}
+\frac{\dis q_iz\phi_i}{\dis 2r_i}\left\{
\ln\left[\frac{\dis \xi_i}{\dis \phi_i}\right]
-\ln\left[\xi_{\rm h}+\sum_{j=1}^M \xi_j \exp(-\beta\epsilon_{ij})\right]
\right\}\right),
\end{equation}
where $\beta=1/(k_{\rm B}T)$, $T$ is the temperature and $k_{\rm B}$ is
Boltzmann's constant.
Denoting the total volume of the lattice by $V=vN_0$, where $v$ is the
volume per site, we obtain the Helmholtz free-energy density $A/VT$
\begin{equation}\label{bgy2}
A/VT=k_{\rm B}\hat{a}/v+a_0 ,
\end{equation}
where $a_0$ is a caloric background which is not discussed further 
in this work. All thermodynamic properties of a mixture can now
be derived from the thermodynamic relation
\begin{equation}\label{bgy3}
d(A/VT)=(U/V)d(1/T)+\sum_{j=1}^M (\mu_j/T)d(\rho_j),
\end{equation}
where $U$ is the internal energy of the system, and where 
$\rho_j=N_j/V=\phi_j/r_jv$ and $\mu_j$ are the number 
density and chemical potential
of component $j$, respectively. The pressure $P$ and the chemical
potentials $\mu_i$, for example, are given by
\begin{equation}\label{bgy4}
P=-\frac{\dis 1}{\dis \beta v}\left[\hat{a}-\sum_{j=1}^M 
\phi_j\left(\frac{\dis\partial \hat{a}}
{\dis\partial \phi_j}\right)_{\beta,\phi_{k\neq j}}
\right] , 
\mbox{ \hspace{3em}} 
\mu_i=\frac{\dis r_i}{\dis \beta}
\left(\frac{\dis\partial \hat{a}}{\dis\partial \phi_i}\right)_{\beta,\phi_{k\neq i}} .
\end{equation}

It now becomes convenient to focus on molar quantities, 
so that $\rho_j$ becomes the molar density of component $j$,
$v$ denotes the volume of a mole of lattice sites,
the energy parameters $\epsilon$ are measured in J/mol,
and $k_{\rm B}$ is replaced by the universal gas constant R.

\section*{Application to one-component fluids}\label{pure}

For the case of a one-component fluid, equations (\ref{bgy1})--(\ref{bgy4})
are evaluated with $M=1$, so that we can drop the subscripts other than h. 
There are three system-dependent parameters which we determine from a 
comparison with experimental data:
the number $r$ of lattice sites occupied by a single 
molecule, the volume $v$ per 
mole of sites, and the interaction energy $\epsilon$.

\begin{figure}[tbp]
\vspace{-7cm}
\centerline{\epsfbox{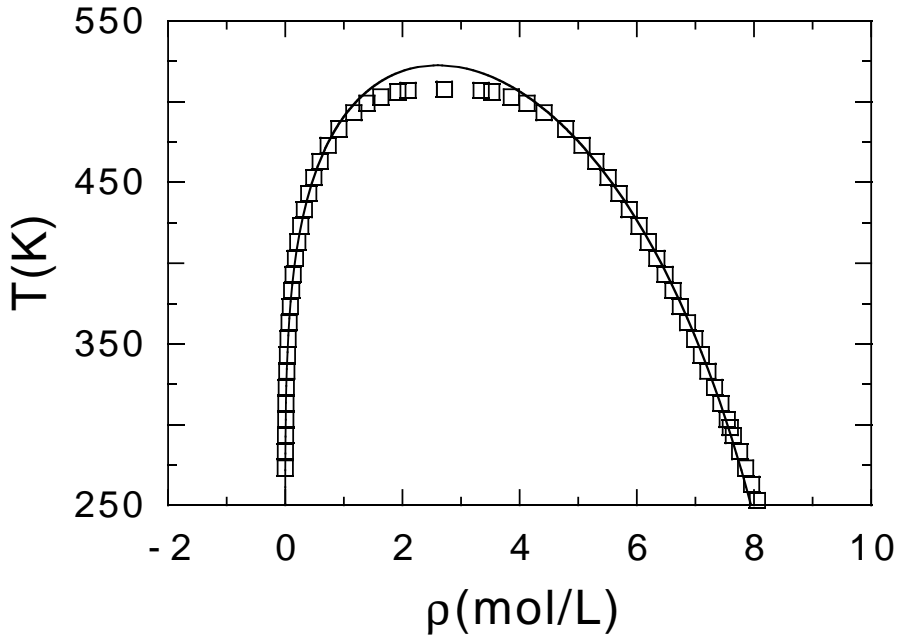}}
\caption{The coexisting vapor and liquid densities for $n$-hexane.
The symbols indicate values from Ref.~\protect{\cite{va75}}, the
solid line values calculated from Eqs.~(\protect{\ref{pure2}})
with system-dependent parameters given in Table~\protect{\ref{tab1}}.
\label{fig1}}
\end{figure}

\begin{figure}[tbp]
\vspace{-9cm}
\centerline{\epsfbox{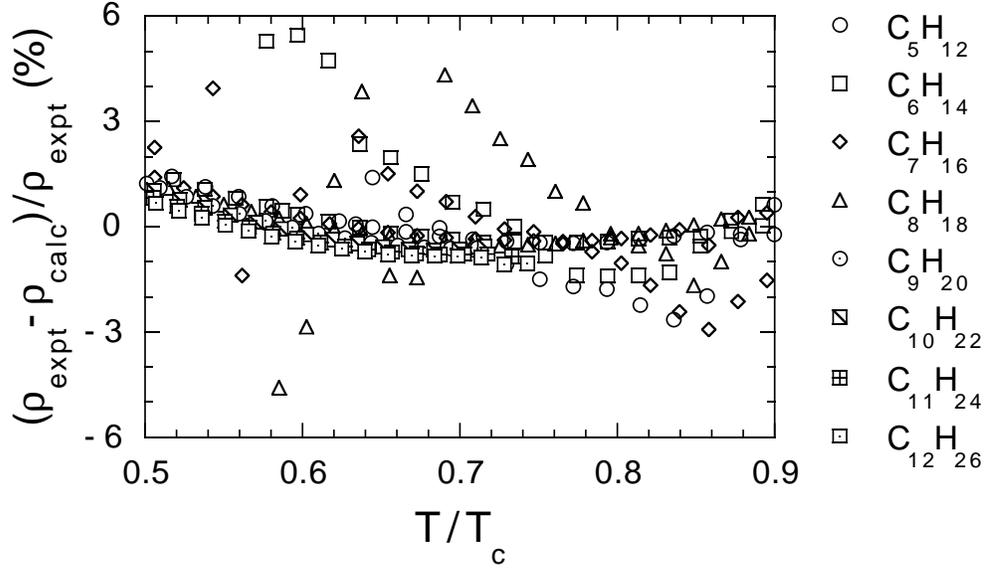}}
\vspace{-7cm}
\caption{Percent deviations between values for the coexistence densities 
as provided in Ref.~\protect{\cite{va75}} and as calculated from 
Eqs.~(\protect{\ref{pure2}}) with system-dependent parameters given
in Table~\protect{\ref{tab1}}.
\label{fig2}}
\end{figure}

To obtain parameters for the $n$-alkanes from pentane through tridecane
we considered coexistence densities. 
The critical temperature $T_{\rm c}$ and  the critical density $\rho_{\rm c}$
of a one-component fluid are determined
by the following two conditions:
\begin{equation}\label{pure1}
\left(\frac{\dis \partial\mu}{\dis \partial\rho}\right)_{T}=0 ,
\mbox{ \hspace{3em}} 
\left(\frac{\dis \partial^2\mu}{\dis \partial\rho^2}\right)_{T}=0 ,
\end{equation}
For temperatures $T$ smaller than $T_{\rm c}$ 
the coexistence densities $\rho_{\rm vap}$ and $\rho_{\rm liq}$ 
can be found from the conditions that pressure and chemical potential
are equal in the vapor and the liquid phase, i.e.\
\begin{equation}\label{pure2}
P(T,\rho_{\rm vap})=P(T,\rho_{\rm liq}) ,
\mbox{ \hspace{3em}} 
\mu(T,\rho_{\rm vap})=\mu(T,\rho_{\rm liq}) .
\end{equation}
We determined the parameters $r$, $v$, and $\epsilon$ for a number
of $n$-alkanes between methane and octane by comparing calculated 
liquid and vapor densities for temperatures between
$0.5T_{\rm c}$ and $0.9T_{\rm c}$ (to avoid the critical region) 
with values tabulated by
Vargaftik~\cite{va75}. We then correlated the parameters
with the chain length $n$ of the alkanes, where we typically excluded the
results for the lowest alkanes. The resulting expressions for the 
parameters as a function of chain length are given by
\begin{eqnarray}\label{pure3}
rv/(\mbox{L/mol})&=&0.016n+0.015 , \\
r\epsilon/(\mbox{J/mol})&=&-2150n-3150 , \label{pure4} \\
\sqrt{r}&=&-0.877539
+1.512447\sqrt{n+1.4651}
\left(\frac{\dis 1}{\dis \sqrt{n}}+0.58259\right) .  \label{pure5}
\end{eqnarray}
The first of these equations, Eq.~(\ref{pure3}), reflects the fact that the
hard core volume of the alkanes is expected to grow linearly with 
chain length, while Eq.~(\ref{pure4}) conveys a linear relationship between
the interaction energy per mole and the chain length $n$. 
Due to end effects, neither relation is expected to hold for very small $n$.
The last equation, Eq.~(\ref{pure5}), is motivated by
the variation of the experimental~\cite{am95} and calculated 
critical temperatures of the alkanes with chain length.
In Table~\ref{tab1} we present the values of the system-dependent parameters
for pentane through tridecane according to 
Eqs.~(\ref{pure3})--(\ref{pure5}). Calculated values for the 
coexistence densities of hexane along with values from Ref.~\cite{va75}
are presented in Fig.~\ref{fig1}, while percent deviations between 
calculated and tabulated values for all alkanes considered in this
work are presented in Fig.~\ref{fig2}. The highest deviations seen
in Fig.~\ref{fig2} are for vapor densities and are, in part, a result
of the correlation of the parameters with chain length. 
The two figures show that our model gives a reasonable description of the 
coexistence densities outside the critical region. 

\begin{table}[tbp]
\vspace{1cm}
\caption{System-dependent parameters for the pure components.
\label{tab1}}
\begin{tabular}{|l|r@{.}l|r@{.}l|r@{.}l|r@{.}l|rl|}
\hline
name &
\multicolumn{2}{r|}{molmass\hspace*{1ex}} &
\multicolumn{2}{r|}{$r$\hspace*{1ex}} &
\multicolumn{2}{r|}{$v$ (mL/mol)\hspace*{1ex}} &
\multicolumn{2}{r|}{$\epsilon$ (J/mol)\hspace*{1ex}} &
LCST(K)& \\
\hline
n-Pentane & 72&146\hspace{1ex} & 9&505229\hspace{1ex} & 
9&994499\hspace{1ex} & -1462&353\hspace{1ex} & 260.4& \\ 
n-Hexane & 86&172 & 10&35107 & 10&72353 & -1550&565 & 371.2& \\
n-Heptane & 100&198 & 11&22030 & 11&31877 & -1622&060 & 435.9& \\
n-Octane & 114&224 & 12&10191 & 11&81632 & -1681&552 & 484.0& \\
n-Nonane & 128&250 & 12&99011 & 12&24009 & -1732&088 & 522.8& \\
n-Decane & 142&276 & 13&88162 & 12&60660 & -1775&729 & 555.5& \\
n-Undecane & 156&302 & 14&77453 & 12&92766 & -1813&933 & 583.7& \\
n-Dodecane & 170&328 & 15&66765 & 13&21193 & -1847&756 & 608.6& \\
n-Tridecane & 184&354 & 16&56028 & 13&46595 & -1877&987 & 630.8& \\
PE  & 140262&0 & 15890&72 & 9&621464 & -1988&320 & & \\ 
PE  & 85000&0 & 9629&90 & 9&621934 & -1988&458 & & \\ 
PE  & 50000&0 & 5663&81 & 9&623040 & -1988&729 & & \\  \hline
\end{tabular}
\vspace{1cm}
\end{table}

\begin{figure}[tbp]
\vspace{-7cm}
\centerline{\epsfbox{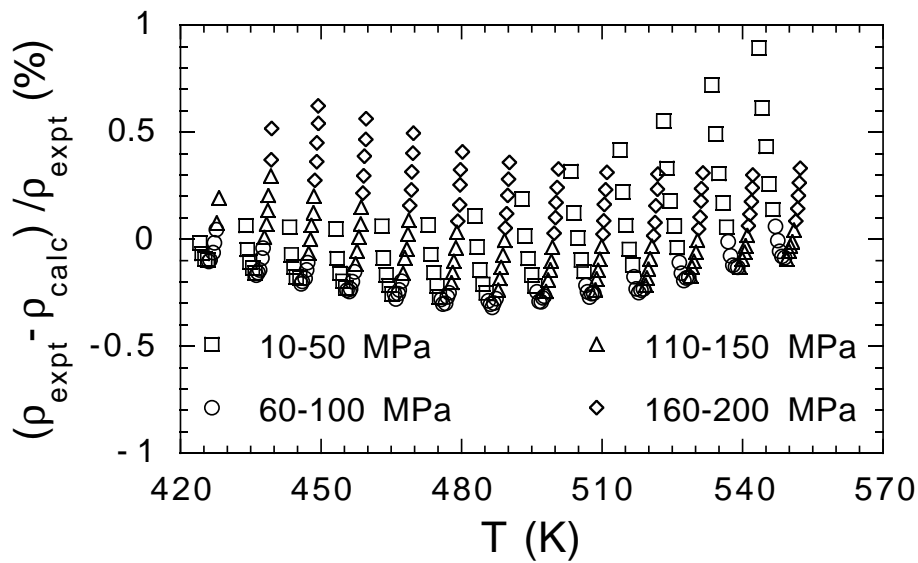}}
\vspace{-6cm}
\caption{
Percent deviations between experimental~\protect{\cite{wa92,kr96}}
and calculated (cf.\ Eq.~(\protect{\ref{bgy4}}))
values of the densities of polyethylene at given temperature and pressure.
\label{fig3}}
\end{figure}

System-dependent parameters for polyethylene (PE) were determined from 
a comparison with experimental PVT data \cite{wa92,kr96} in the 
liquid region for temperatures between 420K and 560K and pressures
between 10MPa and 200MPa. 
When the density of a long chain polymer is given as 
a mass density, rather than a number density, the 
pressure is expected to be independent of the molecular 
mass~\cite{wa92,kr96} of the polymer. 
We determined system-dependent 
parameters for PE from a comparison between
experimental and calculated pressures using three different 
molecular masses for the polymer. The representation of the experimental
data was virtually identical in the three cases. and the 
resulting values for the parameters,
as presented in Table~\ref{tab1}, show that for long
chain polymers $\epsilon$ and $v$ are indeed 
independent of the molecular mass, while $r$ is proportional to it.
A comparison between calculated and experimental values for the density
at given temperature and pressure is presented in Fig~\ref{fig3}. The 
deviations are generally smaller than 0.3\% but reach up to 1\% for
the lowest pressures.

\section*{Phase separation in binary solutions }\label{lcst}

To discuss phase separation in binary mixtures, it is convenient to 
introduce the total molar density $\rho=\rho_1+\rho_2$ and the mole
fraction $x=\rho_2/\rho$ of component 2. 
The thermodynamic field conjugate to the mole fraction $x$ is the difference
of the chemical potentials $\mu=\mu_2-\mu_1$. In terms of these variables, 
the conditions for a point $T_{\rm c}$, $P_{\rm c}$, $x_{\rm c}$ on the 
critical line of a binary mixture take the form~\cite{ro82}:
\begin{eqnarray}\label{lcst2}
\left(\frac{\dis \partial\mu}{\dis \partial x}\right)_{T,P}&=&
\left(\frac{\dis \partial\mu}{\dis \partial x}\right)_{T,\rho}-\rho^2
\left(\frac{\dis \partial\mu}{\dis \partial\rho}\right)_{T,x}^2\left/
\left(\frac{\dis \partial P}{\dis \partial\rho}\right)_{T,x} \right. =0 ,  \\
\left(\frac{\dis \partial^2\mu}{\dis \partial x^2}\right)_{T,P}&=&0 . 
\label{lcst3}
\end{eqnarray}
The first of the two equations defines the spinodal while the second one
implies that, for given pressure, the critical point is an extremum of the 
spinodal in the $T$-$x$ plane. 
Hence,  lower and upper critical solution temperatures (LCST's and UCST's)
correspond to the minimum and maximum of their spinodals, respectively.
For temperatures $LCST<T<UCST$, the mixture phase separates into two
phases of different mole fractions $x_{I}$ and $x_{II}$.
For a given pressure $P$, the coexisting phases  satisfy
\begin{equation}\label{lcst4}
\mu(T,P,x_{I})=\mu(T,P,x_{II}) ,
\mbox{ \hspace{3em}} 
\mu_1(T,P,x_{I})=\mu_1(T,P,x_{II}) .
\end{equation}
For typical solutions of polymers in alkanes, the mole fractions $x$ are very
small numbers so that it is more convenient to employ mass fractions $c$:
\begin{equation}\label{lcst5}
c=M_2x/(M_2x+M_1(1-x)) , 
\end{equation}
where $M_1$ and $M_2$ are the molecular masses of component 1 and 2, 
respectively.

When the BGY lattice model is applied to binary mixtures, 
Eqns.~(\ref{bgy1})--(\ref{bgy4}) are evaluated with $M=2$. 
The total molar density $\rho=\rho_1+\rho_2$ and the mole fraction 
$x=\rho_2/\rho$ of 
the mixture are related to the site fractions $\phi_1$ and
$\phi_2$ by
\begin{equation}\label{lcst1}
\rho=\frac{\dis 1}{\dis v}
\left(\frac{\dis \phi_1}{\dis r_1}+\frac{\dis \phi_2}{\dis r_2}\right) , 
\mbox{ \hspace{3em}} 
\frac{\dis 1}{\dis x}=
1+\frac{\dis \phi_1}{\dis r_1}\frac{\dis r_2}{\dis \phi_2} .
\end{equation}
In addition to the energy parameters $\epsilon_{11}\equiv\epsilon_1$ and
$\epsilon_{22}\equiv\epsilon_2$, and the numbers of segments $r_1$ and 
$r_2$, which are determined directly by the pure components of the mixture, 
values need to be supplied for the interaction energy $\epsilon_{12}$ and 
the volume $v$ of a mole of lattice sites.
For the energy parameter $\epsilon_{12}$ we employ Berthelot's geometrical 
mean combining rule~\cite{ro82}:
\begin{equation}\label{lcst6}
\epsilon_{12}\equiv -\sqrt{\epsilon_1\epsilon_2} . 
\end{equation}
To find a value for $v$, we take advantage of the fact that the BGY model
for the pure polymer depends more strongly on the product $rv$ than
on the values of the number of segments $r$ and the volume $v$ separately.
Hence, for a mixture of an alkane and 
polyethylene, we assign the value $v_1$ of the alkane to the mixture, and 
rescale the segment length of the polymer, i.e.\
\begin{equation}\label{lcst7}
v\equiv v_1 ,
\mbox{ \hspace{3em}} 
r_2\rightarrow (r_2v_2)/v_1 .
\end{equation}
When parameters $v$ and $r_2$, given by Eq.~(\ref{lcst7}), are used to compare 
experimental and calculated $PVT$ data for pure polyethylene over the whole
temperature and pressure range, it is seen that the new parameters diminish
the overall agreement between experiment and theory. However, for the 
lowest pressures the description of the $PVT$ surface is improved over
that using the original parameters, and becomes excellent for the highest
alkanes. This is important since we are going to investigate phase separation
of polyethylene solutions at {\em atmospheric} pressure.

\begin{figure}[tbp]
\vspace{-9cm}
\centerline{\epsfbox{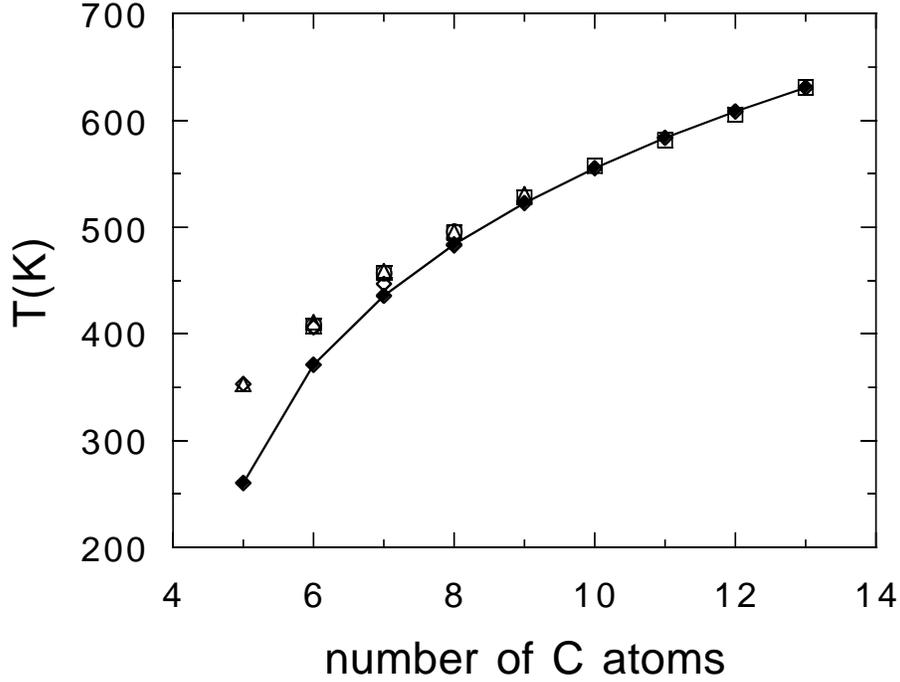}}
\caption{Lower critical solution temperatures for polyethylene in 
$n$-alkanes at $P=0.1$MPa.
The symbols represent experimental data for Marlex-50
by Orwoll and Flory~\protect{\cite{or67b}} (open circles), 
for Kodama and Swinton's~\protect{\cite{ko78}}
longest chain polymer PE Type 2
(open squares), 
extrapolated values to $M\rightarrow\infty$ by 
Hamada et al.~\protect{\cite{ha73}}
(open diamonds), 
and values collected by Charlet and Delmas~\protect{\cite{ch81}} 
(open triangles).
The filled diamonds on the solid line indicate values calculated 
with the aid of Eqns.~(\protect{\ref{lcst1}}) and (\protect{\ref{lcst2}})
and the values presented in Table~\protect{\ref{tab1}} together with
Eqns.~(\protect{\ref{lcst6}}) and (\protect{\ref{lcst7}}).
\label{fig4}}
\end{figure}

To determine the lower critical solution temperatures 
for mixtures of 
polyethylene and the various alkanes discussed in the previous section, 
we employed the parameters presented in Table~\ref{tab1}, 
where the highest molecular-mass parameters were used for polyethylene,
together with Eqs.~(\ref{lcst6}) and (\ref{lcst7}). We then solved
Eqn.~(\ref{lcst2}) numerically to find the spinodal curve for 
atmospheric pressure $P=0.1$MPa. 
The LCST was identified as the minimum of the spinodal curve 
as a function of $x$. 
All calculated LCST's are included in Table~\ref{tab1}, and the
values for the $n$-alkanes between pentane and tridecane
are compared with experimental values~\cite{or67b,ko78,ha73,ch81}
in Fig.~\ref{fig4}. 
These results show that while the BGY lattice model with the 
parameters of Table~\ref{tab1} 
somewhat underestimates the LCST's for the 
shorter chains, it gives a very good description of 
the data for the higher alkanes.

\begin{figure}[tbp]
\vspace{-8cm}
\centerline{\epsfbox{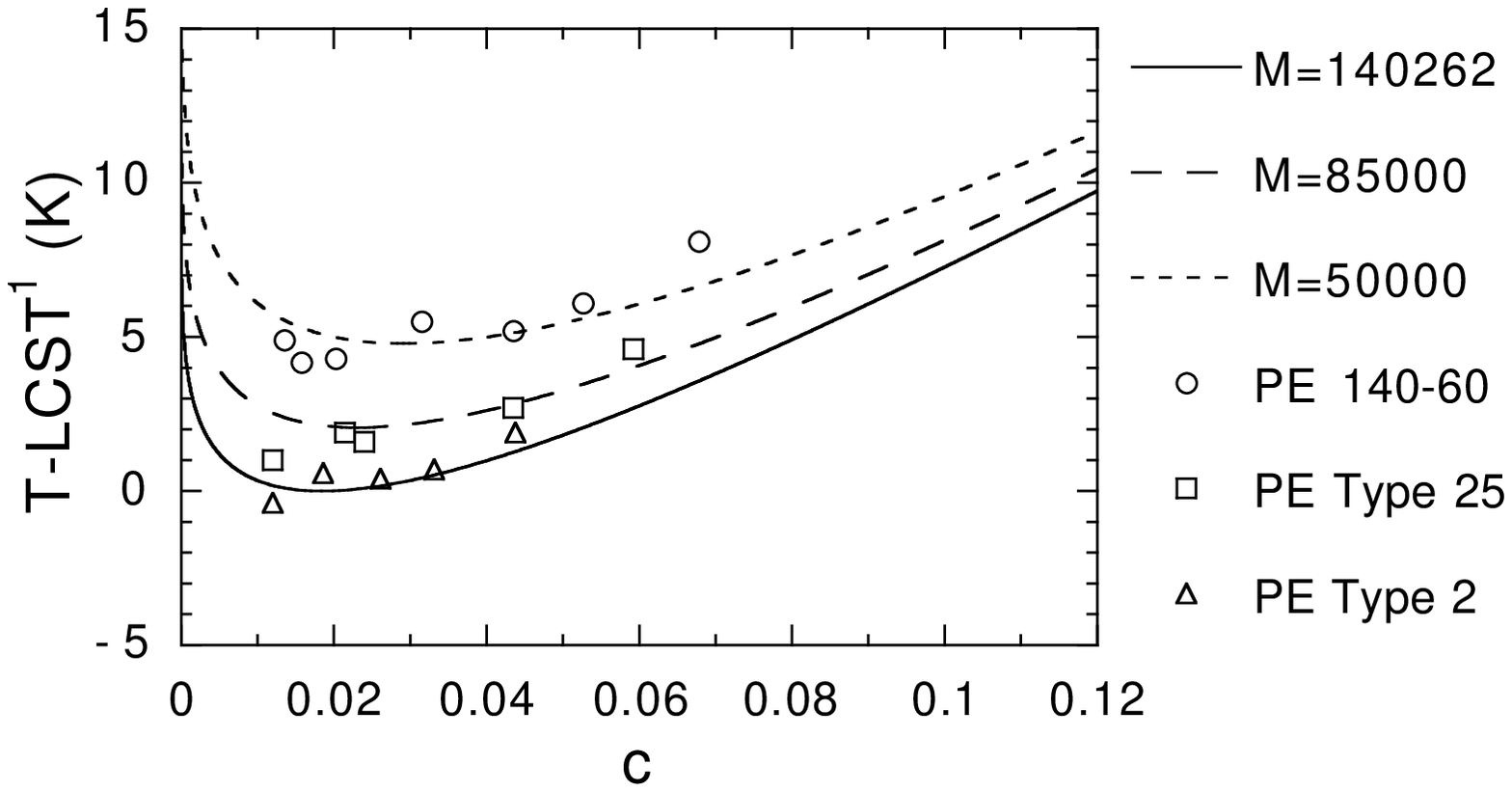}}
\vspace{-8cm}
\caption{Coexistence curves for polyethylene in $n$-decane 
at $P=0.1$MPa as a function of the mass fraction $c$.
The symbols correspond to experimental data by Kodama and 
Swinton~\protect{\cite{ko78}} the lines represent values calculated
according to Eqn.~(\protect{\ref{lcst3}}) with the aid of the 
parameters in Table~\protect{\ref{tab1}} and 
Eqns.~(\protect{\ref{lcst6}}) and (\protect{\ref{lcst7}}).
$^1$~For the experimental data, the LCST ($T=557.6$K) of PE Type 2 has 
been subtracted from the coexistence temperatures, while the predicted LCST 
($T=555.7$K) of PE with molecular mass 140262 has been subtracted
from the calculated values.
\label{fig5}}
\end{figure}

The influence of the chain length of the polymer on its solubility
is illustrated in Fig~\ref{fig5}. The symbols indicate experimental
data  by Kodama and Swinton~\cite{ko78} for three polyethylene 
samples of different molecular masses in decane. We calculated the
coexistence curves for polyethylene in decane for
three comparable molecular masses (see Table~\ref{tab1}) 
using the coexistence conditions Eqs.~(\ref{lcst4}).
Shown in Fig.~\ref{fig5}, as a function of the mass fraction $c$,
are the differences between the coexistence temperatures 
and the calculated and experimental values of the LCST of the 
longest chain, respectively. Since the experimental data in the
paper by Kodama and Swinton~\cite{ko78} were 
not tabulated but presented in a graph 
the comparison is not exact. It appears, however, that the 
effect of changing molecular mass on 
the coexistence temperatures of polyethylene 
in decane is predicted well by the BGY lattice model.

\section*{Conclusion}
With the aid of the system-dependent parameters for the pure components
as well as the geometric mean approximation for the interaction energy,
the BGY lattice model has been shown to predict LCST's for solutions of
polyethylene in alkanes in very good agreement with experimental data.
The theory also captures the effect of changing polymer molecular mass
on solubility. In current work, we are investigating the influence on
miscibility of polymer and solvent architecture.

\section*{Acknowledgements}
The authors would like to thank D.~J.~Lohse for supplying 
$PVT$ data on polyethylene. 
The research is supported by the National Science Foundation through
grant DMR-9424086 and by the Camille and Henry Dreyfus foundation.


\end{document}